\documentclass[useAMS,usenatbib]{aa}
\usepackage{txfonts}
\usepackage{natbib}
\bibpunct{(}{)}{;}{a}{}{,}
\usepackage{longtable}
\usepackage[dvips]{graphicx}
\begin{document}
\title{Using population synthesis of massive stars to study the
interstellar medium near OB associations}
\author{R. Voss\inst{1,2} \and R. Diehl\inst{1} \and D. H. Hartmann\inst{3} \and M. Cervi\~no\inst{4} \and J. S. Vink\inst{5} \and G. Meynet\inst{6} \and M. Limongi\inst{7} \and A. Chieffi\inst{7}}
\institute{Max-Planck-Institut f\"ur extraterrestrische Physik, 
Giessenbachstrasse, D-85748, Garching, Germany
\and Excellence Cluster Universe, Technische Universit\"at M\"unchen, Boltzmannstr.
2, D-85748, Garching, Germany
\and Department of Physics and Astronomy, Clemson University, Kinard Lab of Physics, Clemson, SC 29634-0978
\and Instituto de Astrof\'isica de Andaluc\'ia (CSIC), Camino bajo de Hu\'etor 50, Apdo. 3004, Granada 18080, Spain
\and Armagh Observatory, College Hill, Armagh, BT61 9DG, Northern Ireland, UK
\and Geneva University, Geneva Observatory, CH-1290 Versoix, Switzerland
\and INAF Osservatorio Astronomico di Roma, via Frascati 33, 00040 Monteporzio Catone Roma, Italy
\\
email:rvoss@mpe.mpg.de}
\titlerunning{The interstellar medium around OB associations}
\date{Received ... / Accepted ...}

\offprints{R. Voss}

\abstract{}
{
We study the massive stars in OB associations and their surrounding 
interstellar medium environment, using a population
synthesis code.
}
{
We developed a new population synthesis code for groups of massive stars,
where we model the emission of different forms of energy and matter from the
stars of the association. In particular, the ejection of the two radioactive 
isotopes $^{26}$Al and $^{60}$Fe is followed,
as well as the emission of hydrogen ionizing photons, and the kinetic energy
of the stellar winds and supernova explosions. We investigate various 
alternative astrophysical inputs and the resulting output sensitivities, 
especially effects due to the inclusion of rotation in stellar models.
As the aim of the code is the application
to relatively small populations of massive stars, special care is taken to
address their statistical properties. Our code incorporates both analytical 
statistical methods applicable to small populations, as well as extensive
Monte Carlo simulations.
}
{
We find that the inclusion of rotation in the stellar models has a large
impact on the interactions between OB associations and their surrounding interstellar medium.
The emission of $^{26}$Al in the stellar winds is strongly enhanced, compared to
non-rotating models with the same mass-loss prescription. This compensates the recent
reductions in the estimates of mass-loss rates of massive stars due to the effects of clumping. 
Despite the lower mass-loss rates, the power of the winds is actually enhanced for rotating 
stellar models. The supernova power (kinetic energy of their ejecta) is decreased due to longer 
lifetimes of rotating stars, and therefore the wind power dominates over supernova power for the 
first 6 Myr after a burst of star-formation. For populations typical of nearby star-forming 
regions, the statistical uncertainties are large and clearly non-Gaussian.
}
{}


\maketitle

\section{Introduction}
Massive stars dynamically shape the interstellar medium around them on
timescales of a few Myr \citep[see e.g.][]{Lozinskaya}. Due to their
high mass loss rates \citep{Castor1975,Lamers1976,Barlow} and terminal velocities 
of their winds \citep{Howarth,Lamers1995} and their supernova explosions 
\citep{Jones} they are important sources of mechanical power causing
interstellar medium (ISM) turbulence and the formation of shells and
cavities \citep[e.g.][]{Hucht,Leitherer1992,Maeder1994}.
Their large UV luminosities, they furthermore cause the surrounding
medium to be photoionized to significant distances \citep{Panagia,Vacca}.\\

The mixing of ejecta from young stars into the interstellar medium is
an important process in the interplay between star formation and galaxy 
evolution. A unique window into these processes is provided by the radioactive 
isotope $^{26}$Al, traced by its $\gamma$-ray decay line at
1808.63 keV. With a mean lifetime of $\sim$1 Myr it is a long-term tracer
of nucleosynthesis for populations of sources able to eject it
sufficiently fast after synthesis \citep{PrantzosD}. The COMPTEL instrument
aboard the Compton Gamma Ray Observatory (CGRO) has mapped
the $^{26}$Al emission at 1809 keV in a 9-year full sky survey.
From a comparison between the image morphology and the known spiral
arm tangents and regions of star formation, one can deduce that
massive stars dominate the Galactic $^{26}$Al production, and that
the contribution from novae and AGB stars must be relatively minor 
\citep{Diehl1995,Knodletal1999,Pluschke2001,PrantzosD,Diehl2006}.
Localized groups of massive stars in star-forming regions such as 
Cygnus and Orion emit clear $^{26}$Al signals, and these regions
provide an important contribution to the total amount of $^{26}$Al
present in the Milky Way, and it is therefore important to understand
these basic building blocks. Typically, massive stars eject a few 
$10^{-5}M_{\odot}$ of $^{26}$Al through their winds and supernova (SN) 
explosions \citep[see e.g.][]{Limongi}, and
the total mass of $^{26}$Al in the Milky Way is estimated to be
2.8$\pm$0.8$M_{\odot}$ \citep{Diehl2006}.\\
A similar tracer is the isotope $^{60}$Fe, observed by its 1173 keV and
1333 keV decay lines \citep{Smith2004,Harris2005}. This isotope is
also emitted in the supernova explosions \citep{Limongi}. 
The lifetime of $^{60}$Fe has recently been revised to $\sim$3.8 Myr,
from the previous estimate of $\sim$2 Myr (G. Rugel, private communication).
While the lines are significantly weaker than the $^{26}$Al
line, it is possible to detect them when integrating over large regions
of the Milky Way.\\

We have developed a population synthesis code that follows the
evolution of massive stars and computes the ejection of
$^{26}$Al and $^{60}$Fe from a star-forming region. To facilitate
studies of the dynamics of the surrounding environment, the
code also computes the kinetic energy and the mass ejected
from the winds and the SN explosions, as well as the ionizing flux.
In contrast to {\tt Starburst99} \citep{Leitherer1999,Vazquez2005},
we focus on the study of relatively small, nearby star-forming regions
in the Milky Way. Due to the limited number of massive stars
in thsee individual regions, stochastic processes can cause their
behaviour to deviate strongly from the mean results obtained
for large populations. Because of this, the code is designed
to not only calculate the average behaviour from an
analytical integration over the initial mass function (IMF),
but also the expected distribution of output quantities for finite
populations. This is both done using analytical approximations
and Monte Carlo simulations. To do this we implemented the
statistical methods of \citet{Cervino2006}. These enable us to
determine when Monte Carlo simulations are mandatory in order
to properly describe the distributions, and how many simulations are 
needed to obtain statistically robust results.\\

Here we describe the population synthesis code, and discuss
the dependence on the various physical input models. We compare
our results to the predictions of {\tt Starburst99} for kinetic
luminosity and UV flux, and to \citet{Cervino2000} for radioactive
isotope production. We discuss the simulated stellar populations
in the context of the inter-stellar medium (ISM), with emphasis
on the changes induced by recent stellar evolution tracks for massive 
stars \citep{Limongi,Meynet2005}.

\section{Population synthesis of massive stars}

The synthesis code calculates time profiles of the emission of 
radioactive elements $^{26}$Al and $^{60}$Fe from a population of 
massive stars, for an assumed star formation history. The code also 
traces the energy and mass ejected from the stars through stellar winds
and supernova explosions, as well as the UV radiation they emit.
These quantities are the primary tools necessary to study the
dynamical effects of the interaction between massive stars and
the surrounding ISM. We interpolate between the discrete
inputs from the models, mainly using linear interpolation,
except for temperatures, masses, luminosities, surface gravities 
and stellar ages, which we interpolate logarithmically, unless
otherwise noted.\\

The code is divided into two separate parts: in the first one
stellar tracks are used to calculate time profiles for individual
stars, and we interpolate between these to create a fine grid
of stellar tracks as a function of progenitor mass. For a given 
time we tabulate the stellar properties as a function of initial 
stellar mass, thereby creating isochrones. The second part of the 
code uses the isochrones as input, and for each isochrone calculates
integrated quantities for a given population of stars,
weighting the isochrone values by the IMF. 
This part of the program calculates analytically the mean value of
the distribution of properties we are interested in, as well as some
high order moments that allows an analytical estimation of the probability
distribution of populations with a finite (and small) size
\citep[see][for details]{Cervino2006}. It also estimates these distributions
with Monte Carlo simulations for populations of any size, where intial stellar 
masses are chosen randomly, with relative probabilities according to the 
assumed IMF. These outputs
are complementary in the sense that the average values are
suitable for very large populations of stars, such as entire
galaxies, whereas the Monte Carlo simulations are necessary
for very small star-forming regions. The analytical probability
distributions are useful in intermediate cases ranging from
about one hundred to thousands of massive stars. As we aim
to model star forming regions in the solar neighbourhood,
we restricted our study to stellar models of solar metallicity. 

The details of the physical ingredients
of the code are described in the following. For a number of processes,
switches are included to select between alternative models.
We identify these through convenient parameter names that
are used throughout the paper when discussing the outputs of the code.
The paramer names are summarized in Table \ref{tab:models}.

\subsection{Stellar evolution}
\begin{table*}
\begin{center}
\caption{Parameter names used throughout the paper.}
\label{tab:models}
\begin{tabular}{lll}
\hline\hline
Parameter name & Parameter description & References\\
\hline
{\tt geneva05} & stellar evolution with rotation & \citet{Meynet2005,Palacios}\\
{\tt geneva05alt} & stellar evolution with rotation & \citet{Meynet2005}\\
{\tt geneva97} & stellar evolution & \citet{MaederM1994,Meynet1997}\\
{\tt LC06} & stellar evolution & \citet{Limongi}\\
{\tt wind08} & wind velocities & \citet{Lamers1995,Niedzielski}\\
{\tt wind00} & wind velocities & \citet{Howarth,Prinja}\\
{\tt yieldsLC06} & supernova nucleosynthesis yields & \citet{Limongi}\\
{\tt yieldsWW95} & supernova nucleosynthesis yields & \citet{Woosley1995,Cervino2000}\\
{\tt atmosMS} & stellar atmospheres & \citet{Kurucz,MartinsF,Smith2005}\\
{\tt atmosSmith} & stellar atmospheres & \citet{Kurucz,Smith2005}\\
{\tt atmosOLD} & stellar atmospheres & \citet{Kurucz,Schmutz,Schaerer}\\
\hline
\end{tabular}
\end{center}
\end{table*}

For the stellar evolution of massive stars several alternative sets of
stellar tracks are implemented. The default ({\tt geneva05})
consists of the solar metallicity, rotating stellar models described
in \citet{Meynet2005,Palacios}. These models all have ZAMS
rotation velocities of 300 km s$^{-1}$, producing time averaged equatorial
velocities on the main sequence between 200 and 250 km s$^{-1}$.
There are two versions of these models,
one including the calculation of $^{26}$Al \citep{Palacios} for stars
with initial masses between 25 and 120 $M_{\odot}$, and one 
without \citep{Meynet2005} for stars with initial masses between 9 and
120 $M_{\odot}$. We therefore use the stellar tracks of \citet{Palacios}
above 25$M_{\odot}$, combined with the models of \citet{Meynet2005} below
this limit as our default. However, to enable the calculation of $^{26}$Al,
parts of the numerical calculation was changed between the two sets of
models (G. Meynet, private communication). We therefore test our results
on the stellar energy ejection against those obtained using stellar tracks 
of \citet{Meynet2005} only ({\tt geneva05alt}), see section \ref{sect:energy}.\\

The main effects of rotation on the yields of $^{26}$Al,
and the energy and mass ejection are the following \citep{Meynet2005,Palacios}:
Rotational mixing allows surface enrichment in $^{26}$Al at
an earlier evolutionary stage than obtained in models without rotational
mixing. When rotation is not accounted for, surface $^{26}$Al enrichment
occurs only when the deep layers of the stars where $^{26}$Al is
synthesized are uncovered by the stellar winds. In rotating models
rotational diffusion enables surface enrichments well before the regions
processed by hydrogen burning are uncovered by the stellar winds. This
effect increases the quantity of ejected $^{26}$Al with respect to
non-rotating models. For the same reasons rotating stars present
surface abundance characteristics of the Wolf-Rayet stages before the
stellar winds have uncovered the core. This causes the stars to enter
the WR-phase earlier and to increase the WR lifetimes. For example a
40 (85) $M_{\odot}$ non-rotating star has a WR lifetime of merely
$\sim$0.1 Myr, whereas the rotating counterpart has a WR lifetime
of almost 0.4 (1.4) Myr. Also rotational mixing enables stars with
masses as low as 22$M_{\odot}$ to become WR-stars. These effects
allow rotating models to eject more $^{26}$Al into the ISM than
non-rotating ones. This also increases the mass loss and therefore
the amount of kinetic energy ejected into the ISM. Furthermore
stellar lifetimes are increased by rotation, with a 15-25\% increase in
the hydrogen burning lifetimes. This delays the onset of the
supernova explosions, and lowers the supernova rate for a population
of rotating stars, compared to a population of non-rotating stars.\\

Rotational mixing can in many respects improve the correspondence
between the outputs of stellar models and the observed features of
massive stars. For instance, rotational mixing may be the cause for
surface nitrogen enrichments in OB main sequence stars
\citep[see e.g.][]{Hunter2007,Maeder2008}. With the relatively low
mass loss rates presently favoured (see below), non-rotating models
underpredict the observed ratio of WR to O-type stars. Rotational
mixing favors WR formation, and rotating models are thus in better agreement
with observations. Provided that most of the nitrogen enriched stars
and WR stars are produced by single star evolution, rotation appears
as a key physical ingredient of the models. However, we note that the
comparison of models with observations and the discussion about whether 
the rotating models reproduce the observed properties of massive stars 
better than non-rotating models is currently a subject of debate 
\citep[e.g.][]{Hamann,Meynet2005,Eldridge2006,Vazquez2007}.\\

Two alternative sets of non-rotating models are implemented. One set ({\tt geneva97})
consists of the solar metallicity models of \citet{Meynet1997} combined
with the models of \citet{MaederM1994} for stars below 25 $M_{\odot}$.
The other set ({\tt LC06}) consists of the stellar tracks of \citet{Limongi}.
The main differences between the sets of stellar evolution models
relevant to our study are the inclusion of rotation
in the {\tt geneva05} stellar models, and the updated wind mass-loss
estimates included in the {\tt geneva05} and {\tt LC06} calculations.
These new estimates reduce the wind mass-loss rates of the {\tt geneva05}
and the {\tt LC06} models significantly compared to the {\tt geneva97}
models: in the Wolf-Rayet (WR) phases the inclusion of clumping in the
interpretation of the observational data leads to
reduced mass-loss rates by a factor of $2-3$ compared to the earlier
models \citep{Nugis}. Also the theoretical mass-loss rates
for O and B stars of \citet{Vink2000,Vink2001} are lower by similar factors in
comparison to earlier empirical values of e.g. \citet{Jager1988}.
See the recent review by \citet{Puls} for an extensive description of mass-loss 
rates.\\
We follow each stellar track through the provided evolutionary
points and calculate the quantities needed for the population synthesis.

\subsubsection{Kinetic energy}
\begin{table*}
\begin{center}
\caption{Classification criteria for WR stars, and their wind velocities.
$^{\ast}$$H_s, C_s, N_s$ and $He_s$ are the fractional surface abundances (mass fraction) of hydrogen, carbon, nitrogen and helium, respectively, and the wind velocities are
in km s$^{-1}$.}
\label{tab:WR}
\begin{tabular}{lccc}
\hline\hline
WR type & Surface abundances & Wind velocity ({\tt wind08}) & Wind velocity ({\tt wind00})\\ 
\hline
WNL & $0.4>H_s>0.1$ & 1250 & 1900\\
WNE & $H_s<0.1 \& C_s/N_s<10$ & 2000 & 1650\\
WC6-9 & $H_s<0.1 \& C_s/N_s>10 \& (C_s+O_s)/He_s<0.5$ & 1760 & 1810\\
WC4-5 & $H_s<0.1 \& C_s/N_s>10 \& (C_s+O_s)/He_s<1.0$ & 2650 & 2820\\
WO & $H_s<0.1 \& C_s/N_s>10 \& (C_s+O_s)/He_s>1.0$ & 3000 & 3000\\
\hline
\end{tabular}
\end{center}
\end{table*}
The rate of kinetic energy emitted in the winds of the stars is calculated
from $E_{\mathrm{k}}$=$1/2\dot{M}v_{\infty}^2$, where $\dot{M}$ is the mass-loss
rate from the stellar atmosphere and $v_{\infty}$ is the velocity of the wind at infinity
(the terminal velocity). This represents the energy available once the
wind has escaped the gravitational potential of the star. The interaction of the winds
with the ISM can convert a large fraction of the energy into gas turbulence and radiation.
The mass loss rate is given by the stellar tracks. To calculate the wind
velocity we first coarsely classify the stars according to
the following criteria \citep{Leitherer1999}: Stars with a mass loss 
rate above $\dot{M}=10^{-3.5} M_{\odot}$ yr$^{-1}$ and effective 
temperatures in the range $3.75>\log T_{\mathrm{eff}} <4.4$ are classified as 
luminous blue variables
(LBVs). Stars with an effective temperature above $\log T_\mathrm{eff}=4.0$ and
a fractional abundance of hydrogen at the surface below 0.4, 
are considered WR-stars. They are furthermore divided
into subclasses according to the surface abundances \citep{Smith1991,Leitherer1999}, 
see Table \ref{tab:WR}.\\

The terminal velocity of the wind depends on the type of star.
We utilize two different prescriptions for the calculation: 
in the default mode ({\tt wind08}) stars outside the categories 
defined above are divided into hot stars with 
$v_{\infty}=2.6v_\mathrm{esc}$ and cool stars with $v_{\infty}=1.3v_\mathrm{esc}$ at a
temperature of $\log T_\mathrm{eff}=4.32$ \citep{Lamers1995}, where $v_\mathrm{esc}$ is the
escape velocity at the surface of the star. Luminous blue variables are
assumed to have winds with low velocities (200 km s$^{-1}$) and the
velocities of the WR-winds are estimated from \citet{Niedzielski} and
given in Table \ref{tab:WR}.\\
The alternative mode {\tt wind00} corresponds to the assumptions
in {\tt Starburst99} \citep{Leitherer1992}. Here stars outside the
categories have wind velocities given by 
$v_{\infty}=0.85\times(0.58+2.04\times\log\frac{R}{R_{\odot}})\times v_\mathrm{esc}$ 
\citep{Howarth}, luminous blue variables again have $v_{\infty}=200$ km s$^{-1}$,
and the velocities of the WR-winds are estimated from \citet{Prinja},
and given in Table \ref{tab:WR}.

\subsubsection{Radioactive isotopes}
The wind ejection of the radioactive isotope $^{26}$Al is followed along
the stellar tracks. The ejection rate is found from multiplying the
surface abundance fraction of $^{26}$Al with the mass loss rate,
and the amount of $^{26}$Al present in the interstellar medium around
the star is calculated by taking into account the decay timescale of
$\sim$1 Myr. The other radioactive isotope considered in this study,
$^{60}$Fe, is not ejected in the stellar winds, but only in the supernova
explosions. For this isotope we consider two decay timescales, with $\sim$2 Myr as
the default and $\sim$3.8 Myr suggested by recent measurements. 
At the last point of the stellar tracks the radioactive yields from
the supernova explosions are added. We have two prescriptions for
calculating the yields: in the default mode ({\tt yieldsLC2006}) we
use the yields calculated by \citet{Limongi}. The yields are found
based on the initial mass of the stars, interpolating linearly between
their data points. In the alternative method ({\tt yieldsWW95}) the
yields below initial masses of $25M_{\odot}$ are taken from 
\citet{WoosleyW1995}. Above this mass we set the output of each star
to $5.0\cdot10^{-5}M_{\odot}$ for both $^{26}$Al and $^{60}$Fe. This is
approximately the average yields found in \citet{Cervino2000} from linking
the core mass of the {\tt geneva97} stellar tracks with the supernova
yields of \citet{Woosley1995}. We note that in \citet{Cervino2000},
the results depend strongly on the method used (see their Figs. 1 and 2).
Our crude approximation therefore reflects unknown physics. 
The total yield is not affected dramatically
by these uncertainties, as there are few massive stars, and for 
$^{26}$Al, the emission from these stars is dominated by the wind
contribution.

\subsection{Creation of the stellar isochrones}
We use the stellar tracks to create isochrones for a user-defined
set of times after the onset of star formation. To facilitate the
interpolation between the stellar tracks, for each of the tracks
a subset of 51 equivalent evolutionary points are extracted according to the
definitions given in \citet{Maeder1990}. The reduction of
the stellar tracks to 51 points eliminates some of the finer details of 
stellar evolution, but the uncertainties due to this simplification are small
compared to the uncertainties of defining equivalent evolutionary
points and interpolating between stellar models with different initial mass.   
For each isochrone an array of
initial stellar masses is defined, and stellar tracks are produced by
interpolation between the equivalent points of the nearest input stellar
tracks. We note that the ages at the equivalent points are interpolated
using a cubic spline in logarithmic space, as lower degree interpolation
methods create artificial results 
\citep[e.g. in the supernova rate, see][Fig. 1]{Cervino2001}.
The isochrone values are then found by interpolation along these
stellar tracks. We note that the {\tt geneva05} tracks are only calculated
to the end of central helium burning, while the {\tt geneva97} tracks
are followed to the end of central carbon burning. We use the last
available grid points and extrapolate it towards the end, thus
neglecting the stages
between the ends of the stellar tracks and the supernova explosions.

\subsection{Calculation of population properties}
The stellar isochrones are the basis of our population synthesis.
Each isochrone is used to calculate the population properties, both
by integrating the isochrone analytically and using Monte Carlo techniques.
Both methods sample the number of stars of a given mass according
to the initial mass function (IMF), for which
we use the Salpeter mass function \citep{Salpeter} as the default,
with  8$M_{\odot}$ and 120$M_{\odot}$ as lower and upper integration
limits. As only massive stars are
investigated here, we do not consider more complicated IMF models
\citep[e.g.][]{Scalo,Kroupa} that also address the shape of the mass 
function at lower masses. We furthermore include
an implementation of the statistical methods described 
in \citet{Cervino2006} to account for statistical variability of
stellar-mass weightings for smaller numbers of stars. This
allows us to calculate analytically the statistical distribution of 
the output quantities, depending on the number of stars in the stellar
population, similar to what can be done with Monte Carlo simulations.
We do not include the effect of binaries \citep{Vanbeveren2007,Eldridge2008}. 
The frequency of interacting binaries,
and the effects of the interactions are poorly known, and the effects
on the parameters studied in this paper are difficult to assess. Such
interactions might remove outer layers of primary stars, and rejuvenate
secondary stars. Both processes can potentially enhance the $^{26}$Al
ejection \citep{Langer1998}.


\subsubsection{Supernova explosions}
In this study, we assume that the kinetic energy ejected in a supernova
explosion is $10^{51}$ erg s$^{-1}$ \citep[e.g.][]{Woosley1995}, irrespective 
of the progenitor mass. The supernova rate is derived as an average 
over a period of 0.1 Myr and from this the 'kinetic luminosity' of the 
supernovae is calculated. The mass ejected in supernova explosions
is found from the difference between the mass of the last point in the 
stellar evolution code, and the mass of the compact remnant. We assume that 
all neutron stars have a canonical mass of 1.4$M_{\odot}$ and that all
black holes have a mass of 7.0$M_{\odot}$. The supernova explosions of
stars with initial masses below 25$M_{\odot}$ are assumed to produce neutron 
stars whereas stars above this mass limit are assumed to produce black holes. 
Our results are not sensitive to these values within reasonable ranges. For
example, if the most massive stars end up as neutron stars due to
mass loss, this will only change the mass of the ejected matter by
$\sim5-10$\%, as the majority of the mass is ejected in the winds
at earlier evolutionary stages.  It should 
be noted that the {\tt geneva97} and {\tt geneva05} stellar
tracks do not cover the final stages of stellar evolution, and the mass
ejected in the supernova explosions may therefore be overestimated. However,
the mass loss from stellar winds is underpredicted by the same amount,
and these effects should cancel out to some extent.\\

\subsubsection{Ionizing flux}
The ionizing flux of each star is calculated by linking stellar
atmosphere models to the state of the star. The surface gravity
and the effective temperature are used to find the atmosphere model
with the closest resemblance. In the WR-phases, we use the method of
\citet{Smith2005} to find these values at an optical depth of $\tau=10$. 
In this method, the effective temperature is simply given by 
$T_\ast=0.6\cdot T_\mathrm{hyd}+0.4\cdot T_{2/3}$, where $T_\mathrm{hyd}$ 
is the hydrostatic effective temperature of the stellar models, and
$T_{2/3}$ is the corrected surface temperature given in the
{\tt geneva97} and {\tt geneva05} stellar models. This parameter
is unavailable in the {\tt L\&C05} stellar models, and the ionizing
flux is therefore not calculated for these.\\
The default set of atmosphere models ({\tt atmosMS}) applied for OB stars 
\citep{MartinsF} and for WR stars \citep{Smith2005} both take non-LTE, wind and
line-blanketing effects into account. For stars with parameters beyond 
the range of these models we use \citet{Kurucz} atmospheres. 
For comparison with {\tt Starburst99} we include an option of using
the models of \citep{Smith2005} for both WR and OB stars ({\tt atmosSmith}),
and for comparison with earlier studies an alternative set of models
({\tt atmosOLD}) consisting of the WR models of \citet{Schmutz} and
the OB models of \citet{Schaerer}, again supplemented by the models
of \citet{Kurucz}.

\begin{figure}
\resizebox{\hsize}{!}{\includegraphics[angle=0]{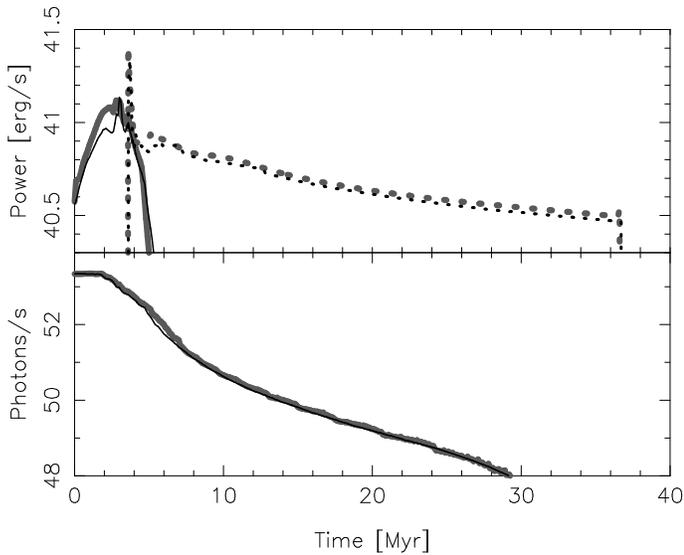}}
\caption{Comparison between {\tt Starburst99} (thin black lines) 
and our results (thick grey lines) for a co-eval population. 
top: time profiles of the kinetic 
power emitted through the stellar winds (solid) and supernova
explosions (dotted). Bottom: The ionizing flux.
}
\label{fig:starburst99}
\end{figure}

\section{Population synthesis results}
We combine the results obtained from integrating over the individual
isochrones to create time profiles of the average ejection of $^{26}$Al, $^{60}$Fe,
kinetic energy, hot gas and the ionizing flux. In this section we discuss
these individual quantities, and their dependence on the assumptions made.

\subsection{Comparison with previous studies}
Synthesis of stellar populations is an important tool for
understanding stellar clusters and galaxies. Since the first studies 
\citep{Tinsley,Bruzual83} the sophistication of the models have increased
drastically \citep[e.g.][]{Bruzual2003}. The majority of these models
are designed to predict the spectra of large stellar populations over
a wide range of ages and metallicities, for which the exact parameters
of the few massive stars are not of major importance. This is quite 
different from the goal of our study, which is the calculation of 
parameters of interest for the interactions between massive stars and 
their environment. Most studies are therefore not suited for comparison
with the present study. Thus we use {\tt Starburst99} for comparing
our results on the ionizing flux and kinetic luminosity. For this purpose
we produce a model with parameters similar to parameters of their studies.
We choose a model with a Salpeter IMF \citep{Salpeter}, between
$8-120 M_{\odot}$ with the parameters {\tt geneva97}\footnote{
While the {\tt geneva05alt} stellar models are included in {\tt Starburst99}
\citep{Vazquez2007}, this part of the code is not publicly available yet.}, 
{\tt wind00},
{\tt atmosSmith} and {\tt yieldsWW95}. The comparison 
for a cluster with a total mass of 10$^6 M_{\odot}$ in the given mass 
range are shown in Fig. \ref{fig:starburst99}.
There is obviously good agreement between both models. The few discrepancies
(e.g. the somewhat larger wind power between 2 and 3 Myr in our model)
can all be traced back to different interpolation schemes, and
therefore represent unknown physics due to the limited stellar grid. The only
previously published population synthesis of $^{26}$Al and $^{60}$Fe
is \citet{Cervino2000}. Again we find good agreement with their
results. Only the early supernova ejection of $^{26}$Al is higher in our
model. This is due to the fact that their supernova yields decrease for
the highest mass stars, whereas for our parameter {\tt yieldsWW95} 
they are constant above 25$M_{\odot}$. 

\begin{figure}
\resizebox{\hsize}{!}{\includegraphics[angle=0]{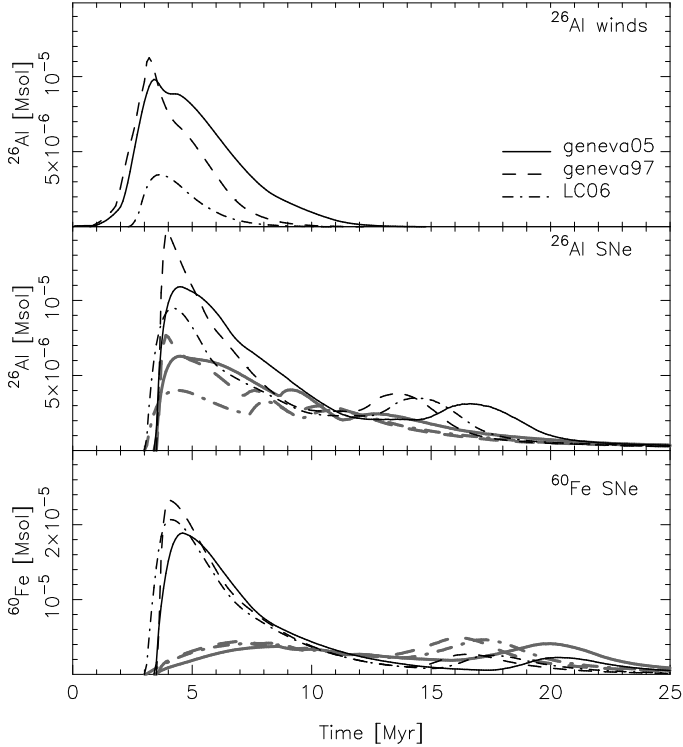}}
\caption{Average time profiles of $^{26}$Al and $^{60}$Fe for a coeval
population of stars, for the three different sets of stellar models 
available. The lines show the amount of the elements present in the
ISM per star in the $8-120M_{\odot}$ range.
The $^{26}$Al yields are divided between the wind and the
supernova contributions, whereas the wind contribution of $^{60}$Fe
is negligible. The black lines indicate the results using the
{\tt yieldsLC2006} supernova yields, whereas the grey lines indicates 
the results using the {\tt yieldsWW95} yields.
}
\label{fig:compare_yields}
\end{figure}

\subsection{Radioactive isotopes}

\begin{figure}
\resizebox{\hsize}{!}{\includegraphics[angle=0]{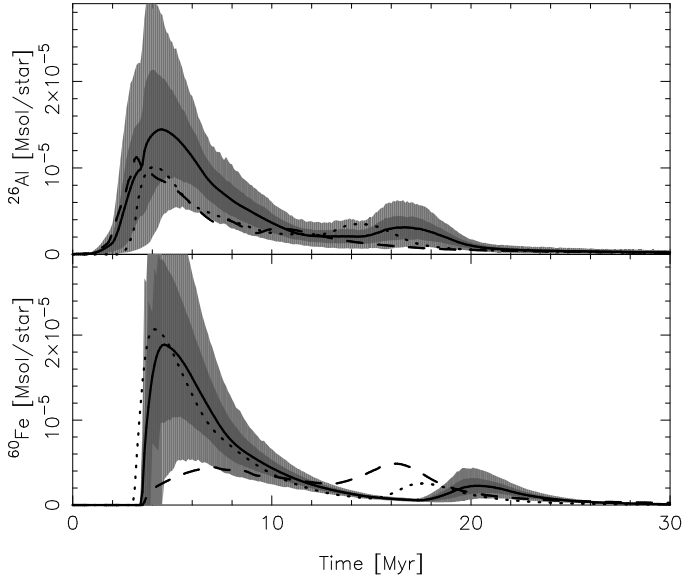}}
\caption{Time profiles of $^{26}$Al (top) and $^{60}$Fe (bottom) for a
coeval population of stars. The solid 
lines indicate the average profiles for the {\tt geneva05} stellar
models with the {\tt yieldsLC2006} supernova yields. The
the dark and light grey regions show the 1$\sigma$ and the 2$\sigma$
deviations for a population of 100 stars between 8 and 120 $M_{\odot}$,
based on Monte Carlo simulations. The dashed and dotted lines show
our main alternative models. Dashed: {\tt geneva97} stellar tracks
with {\tt yieldsWW95} supernova yields. Dotted: {\tt LC06}
stellar tracks with {\tt yieldsLC2006} supernova yields. }
\label{fig:al_timeprof}
\end{figure}

\begin{figure}
\resizebox{\hsize}{!}{\includegraphics[angle=0]{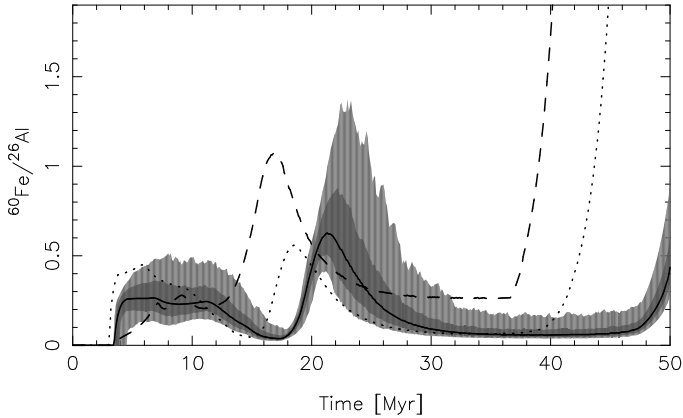}}
\caption{Time profiles of $^{60}$Fe/$^{26}$Al emission ratio for a
coeval population of stars. Legends same as Fig. \ref{fig:al_timeprof}.}
\label{fig:ratio_timeprof}
\end{figure}

\begin{figure}
\resizebox{\hsize}{!}{\includegraphics[angle=0]{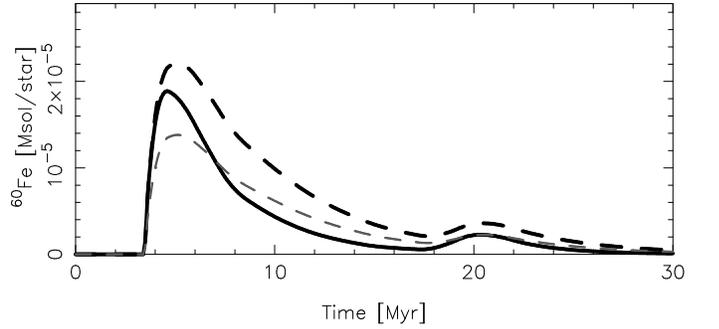}}
\caption{Time profiles of $^{60}$Fe for a coeval population of stars
for the two different decay timescales. The solid line is the same as 
in Fig. \ref{fig:al_timeprof}, with an average lifetime of $\sim2$ Myr,
whereas the black dashed line is for an average lifetime of $\sim$3.8 Myr.
The grey dashed line indicates the observational difference (the amount
of $^{60}$Fe that would be inferred from observations if a lifetime of
2 Myr is assumed, in the case the lifetime actually is 3.8 Myr.}
\label{fig:fe60_half}
\end{figure}

Compared to the previous population synthesis study of 
$^{26}$Al and $^{60}$Fe \citep{Cervino2000}, our study includes new
stellar models, specifically {\tt geneva05} and {\tt LC06}. 
Both take the reduced mass loss estimates into account,
and the {\tt geneva05} model furthermore includes the effects of stellar
rotation. The {\tt LC06} model consistently includes
the supernova explosions of the very massive stars.
While these improvements have been discussed for individual stars and for the
overall ejection from a population of stars \citep{Palacios,Limongi}, 
the effects on the time profile have not been evaluated in detail.
The time profiles of $^{26}$Al and $^{60}$Fe in the interstellar
medium for a coeval population of stars are shown in Fig. 
\ref{fig:compare_yields}. The $^{26}$Al has been divided into the wind
and supernova contributions. Only the supernova contribution
to $^{60}$Fe is shown, as the wind contribution is negligible.\\
The wind contribution of $^{26}$Al is relatively similar for the
{\tt geneva05} and the {\tt geneva97} models. This is quite surprising,
given the relatively large differences between these models. The reason
is that the lower wind mass-loss in the {\tt geneva05} models is
compensated by two effects. One is the rotational mixing of elements,
which leads to larger surface abundances. The other is the longer lifetime
of the rotating models, which causes the amount of material lost to be
similar to the amount lost in the {\tt geneva97} model, despite the lower
rate of mass loss. This is also the reason that the $^{26}$Al time
profile shows a more extended peak for the rotating models.
The {\tt LC06} models, on the other hand, predict much less $^{26}$Al
ejection by the pre-supernova wind, for two main reasons: with the 
reduced mass loss, the rate of ejection is low, and
the $^{26}$Al is not effectively mixed to the surface.\\
For the supernova ejections the differences between the model 
predictions of $^{26}$Al and $^{60}$Fe are smaller, with the main effect
coming from the longer lifetime of the rotating models causing a
slightly less peaked profile. On the other hand, drastic differences
between the predictions for the first 10 Myr are obvious when comparing
{\tt yieldsWW95} and {\tt yieldsLC2006}. This
probably mainly reflects the uncertainties in the final structure
of the very massive stars, at the time of the supernova explosions.
The low mass-loss in the models of \citep{Limongi} gives very
massive cores, that produce large amounts of the elements. 
The higher mass-loss rates assumed in the study of \citet{Cervino2000}
gives lighter cores, and this results in the {\tt yieldsWW95}
being lower than the {\tt yieldsLC2006}.
We note that with the very large theoretical uncertainties, observations of 
$^{26}$Al and $^{60}$Fe thus have the
potential to place interesting constraints on the final evolutionary
state of very heavy stars, although in an indirect way, since only the
integrated effect of different sources can be observed.\\
In Fig. \ref{fig:al_timeprof}, the average time profiles of $^{26}$Al 
(including both the wind and the supernova contributions) and 
$^{60}$Fe are shown for different models, together with the statistical
variance. Shown are the 1$\sigma$
and 2$\sigma$ statistical deviations (the intervals containing
68\% and 95\% of the Monte Carlo simulations) for a population of 100 stars
in the $8-120M_{\odot}$ range\footnote{1 star in this range corresponds
to 381, 140 and 188 stars in the $0.1-120 M_{\odot}$ range for the
Salpeter \citep{Salpeter}, Kroupa \citep{Kroupa} and the
Scalo \citep{Scalo} mass functions, respectively, and to 13\%, 18\%
and 5\% of the stellar mass.}, corresponding to a typical nearby
star forming region \citep[for example the number of massive stars
formed within the last 15 Myr in the Orion OB1 association is estimated
to be close to 100][]{Brown1994}. We note that the 8--120$M_{\odot}$
range includes more massive stars than are observed in many nearby
regions. However, when a probabilistic description 
(such as our Monte Carlo simulations) is assumed, 
the limit should be the most massive star theoretically possible in the cluster.
Observed clusters correspond to random realizations of the IMF and
the most massive stars in these can therefore have much lower masses
than the upper limit of 120$M_{\odot}$ (see also section \ref{sect:discussion}
where the contributions of various initial mass ranges to the observables
are shown).\\
From Fig. \ref{fig:al_timeprof} it is obvious that for relatively small 
populations, it is essential that these statistical effects are 
taken into account, when interpreting observations. Also very interesting
is the ratio between the observable strengths of the $^{60}$Fe
and $^{26}$Al $\gamma$-ray lines. This is shown in Fig. 
\ref{fig:ratio_timeprof} for the same models as in Fig. \ref{fig:al_timeprof}.
For much of the time, this ratio places stronger
constraints on the stellar models than the individual observations of
 $^{26}$Al and $^{60}$Fe. This is due to the fact that the emissions of
these two elements are correlated. The strong increase in the 
$^{60}$Fe/$^{26}$Al ratio seen around the lifetime of an 8$M_{\odot}$ star
($\sim35-50$ Myr, depending on the stellar model) is simply an effect
of the longer lifetime of $^{60}$Fe, when the elements are not being
replenished (non-steady state situation).\\
Recent results (G. Rugel, in preparation) indicate 
that the lifetime of $^{60}$Fe is significantly ($\sim$3.8 Myr) 
longer than the commonly used lifetime of $\sim2$ Myr. It is unclear if
a different lifetime would have any significant impact on the nucleosynthesis 
of $^{60}$Fe, and a study of this is
beyond the scope of this paper. In Fig. \ref{fig:fe60_half} we show
the effect on the time profile of $^{60}$Fe for our default model,
assuming that the amount of $^{60}$Fe released in the supernova explosions is
unchanged. Due to the longer decay timescale, the build-up of the
isotope is larger, and in the period after the main peak ($5-15$ Myr
after the star formation), the amount of $^{60}$Fe present in the
ISM is approximately twice as large as for the shorter $^{60}$Fe lifetime.
Note that the effect on the observed flux
is different. While there is more $^{60}$Fe present in the ISM, the
$\gamma$-ray emission per unit mass is decreased due to the longer
decay timescale. This is illustrated by the grey dashed line in
Fig. \ref{fig:fe60_half}: the integrated amount of $\gamma$-ray emission
is unchanged (since the amount of $^{60}$Fe released
from the stars is unchanged), but the distribution is slightly 
shifted to later times. The only significant effect is the lower peak
at $\sim5$ Myr. It should be noted that in a constant star-formation
scenario (steady-state), 
the measured ratio of $^{60}$Fe/$^{26}$Al {\it emission} is
independent of the decay times of the two isotopes (although
the ratio of $^{60}$Fe/$^{26}$Al does depend on the decay times).

\subsection{Energy and mass}
\label{sect:energy}
\begin{figure}
\resizebox{\hsize}{!}{\includegraphics[angle=0]{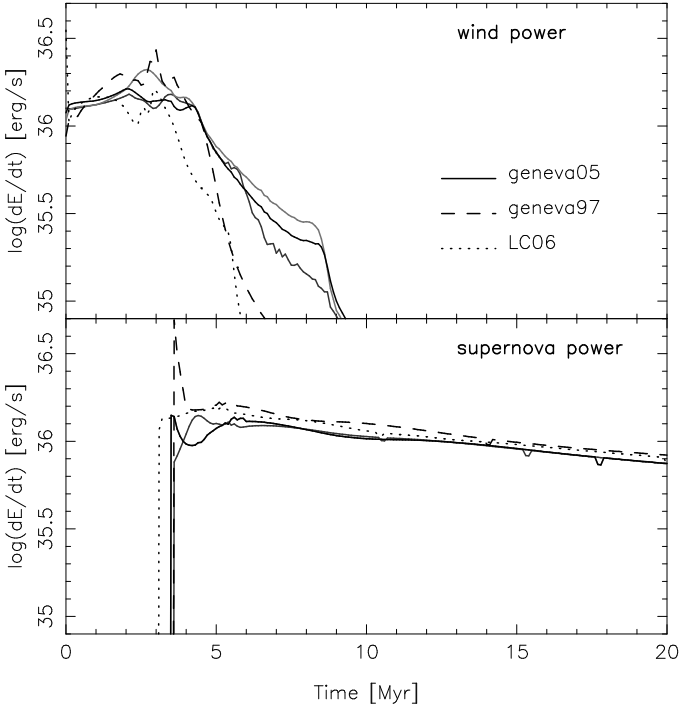}}
\caption{Time profiles of the wind (top) and supernova power per star
from a coeval population of stars in the $8-120M_{\odot}$ mass range,
using the various stellar models with the {\tt wind08} wind velocities.
The solid dark grey line is associated with the {\tt geneva05alt}
evolutionary tracks, and the solid light grey line corresponds
to the {\tt geneva05} stellar models with {\tt wind00}.}
\label{fig:compare_energy}
\end{figure}

The energy and mass ejection rates are the most important parameters
for the evolution of the interstellar medium in the vicinity of young 
stars. Winds and SN ejecta compress the surrounding medium,
creating wind-driven bubbles, supernova remnants, and even super-bubbles
in the case of multiplets of massive stars. These stellar systems dissociate 
the surrounding molecular clouds and may induce star-formation within these.
In Fig. \ref{fig:compare_energy}, the "kinetic luminosities" of the
winds and the supernova explosions are shown for the three sets of
stellar tracks. In the first 3-4 Myr, no
stars explode, and the kinetic luminosity originates exclusively from
stellar winds. The mechanicial power of these sustained winds is roughly
equal to the average power of punctuated supernova explosions, once these 
set in. It is important to note that integrated over the first 10 Myr,
the total energy emitted in winds is actually larger than the energy 
associated with the supernova explosions. This is important for shaping
the medium around a star-forming region. There are large differences
between the various stellar models. In particular it is worth noting
that while decreased mass-loss rates in rotating stellar models obviously 
decrease the wind power, this is compensated for by their longer lifetimes. 
Furthermore the longer stellar lifetimes lower the
supernova rate (except at late times $\gtrsim$35 Myr, where the supernova
rate for non-rotating stars becomes zero, whereas the rotating stars still
explode until $\sim$45 Myr), and therefore the early dominance of wind
energy is even stronger.\\ 
A comparison between the two wind prescriptions {\tt wind00} and {\tt wind08} 
shows significant differences. Wind velocities of different types
of massive stars still have very large uncertainties, and this translates 
into large uncertainties in the implicated wind power.
As the most massive stars explode after $\sim5$ Myr, the
wind power decreases rapidly, thereafter the supernova rate decreases steadily
until stars with masses $\sim8M_{\odot}$ explode, and
the energy deposition in the ISM becomes negligible.\\

\begin{figure}
\resizebox{\hsize}{!}{\includegraphics[angle=0]{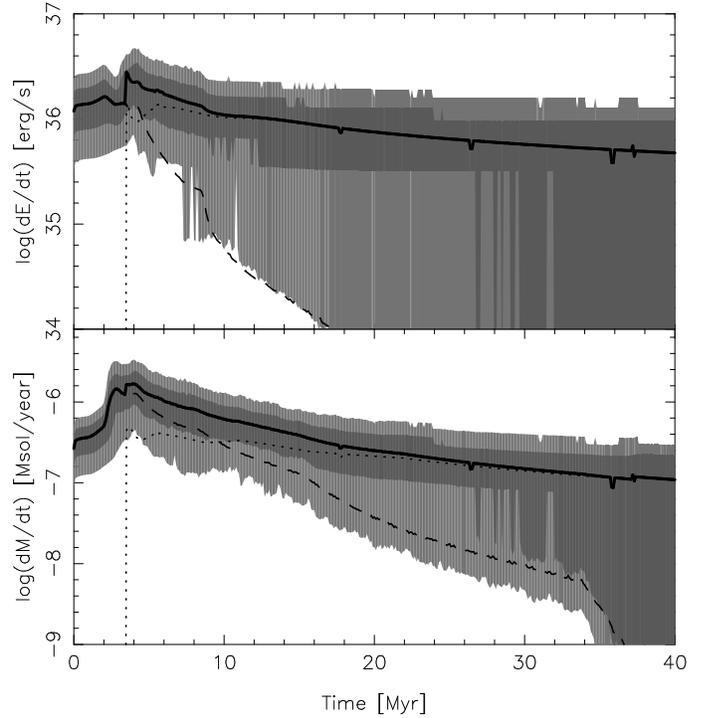}}
\caption{Time profiles energy (top) and mass (bottom) ejection rates, with
1$\sigma$ (dark grey) and 2$\sigma$ (light grey) Monte Carlo deviations
for a population of 100 stars between 8 and 120 $M_{\odot}$. The solid
line indicates the full outputs, with the dashed and dotted lines
indicating the contributions from stellar winds and supernova explosions,
respectively. The Monte Carlo outputs were averaged over 1 Myr.}
\label{fig:energy}
\end{figure}

In Fig. \ref{fig:energy} the average total kinetic luminosity and
the mass ejection rate are compared through Monte Carlo simulations of a
small population of stars. For the Monte Carlo simulations the power
was averaged over 1 Myr, and still the variations are very large,
especially in the later stages, where only few supernovae are likely
to explode within a time interval. When studying small regions it
is important to note these statistical effects, especially for processes
with shorter timescales, where the variability of a local supernova rate
increases drastically. For example the 1$\sigma$ range of the
supernova rate around 10 Myr covers more than a factor of 100 when
averaged over a timescale of 0.1 Myr, similar to the replenishment timescale 
of the X-ray emitting gas in the region of Orion studied by \citet{Gudel}. 
We note that the variation in the time integrated energy 
\citep[studied by][]{Cervino2001,Cervino2002}, important for studying
the X-ray bubble expansion at larger scales, is smaller, since it
only depends on the number of supernova explosions and not on the specific
explosion times.
The mass ejection rate shown in the lower panel of Fig. \ref{fig:energy}
is more strongly dominated by the ejection through stellar winds,
and as this is a continuous process, the variations are relatively smaller.

\subsection{UV radiation}
\begin{figure}
\resizebox{\hsize}{!}{\includegraphics[angle=0]{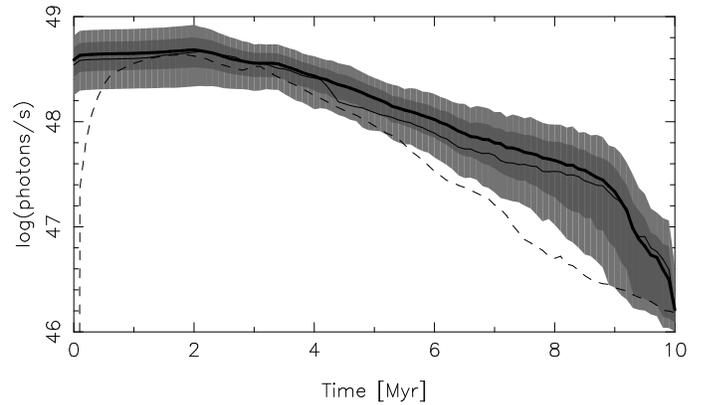}}
\caption{Time profiles of the ionizing flux per star, 
with 1$\sigma$ (dark grey) and 2$\sigma$ (light grey) Monte Carlo 
deviations
for a population of 100 stars between 8 and 120 $M_{\odot}$, using
the {\tt geneva05} stellar models with the {\tt atmosMS} atmosphere
models. The thin solid line was produced by using the {\tt atmosOLD}
atmosphere models, whereas the dashed line uses the {\tt LC06} stellar
models and the {\tt atmosMS} atmospheres.}
\label{fig:UV}
\end{figure}

The bulk of the ionizing flux is provided by hot stars that disappear
after only a few Myr. This is clearly seen in Fig. \ref{fig:UV}, where
the time profile of the ionizing flux is shown. The flux is high and
roughly constant during the first 2 Myr, after which it declines.
After less than 10 Myr it has declined by a factor of more than 100, and
the rate of decline steepens further. There are still large uncertainties in the
modelling of non-LTE expanding atmospheres. Comparing the results with a
now superseded set of stellar atmospheres shows only minor differences, but the
comparison with the non-rotating {\tt L\&C} stellar models shows dramatic
differences. So while the correct atmosphere modelling is important for 
the study of single stars and for understanding the UV spectrum, 
the stellar evolution modelling is much more important for the 
understanding of the interactions between groups
of massive stars and their surroundings. It is important to note that
for small stellar associations, the statistical deviations from the mean are
very large \citep[see e.g. Fig. 4 in][for the deviation for associations of
different masses]{Cervino2000b}.

\subsection{Statistical properties of the stellar populations}
\begin{figure}
\resizebox{\hsize}{!}{\includegraphics[angle=0]{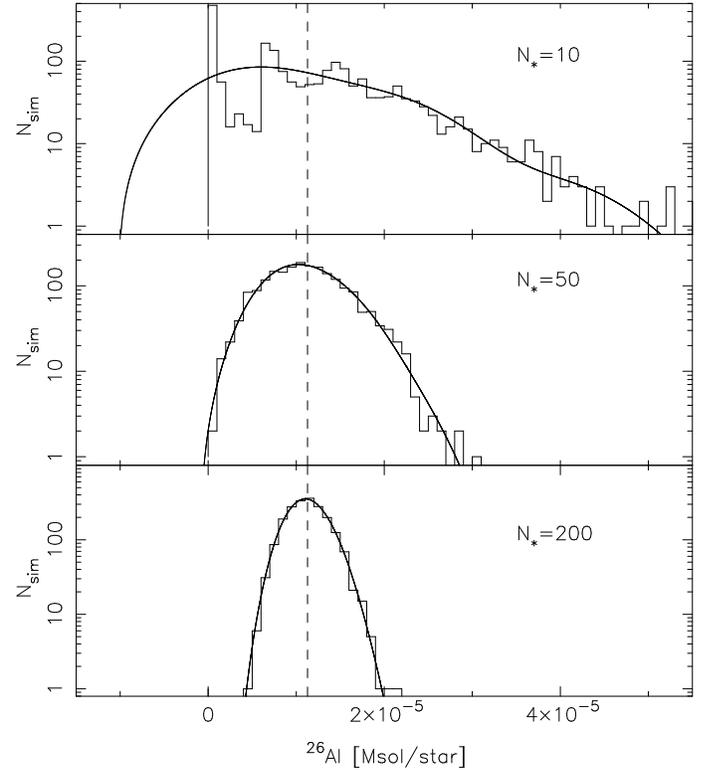}}
\caption{The statistical distribution of $^{26}$Al in the interstellar
medium around a cluster of 10, 50 and 200 massive stars ($N_{\ast}$) 
in the $8-120 M_{\odot}$ range in the top, middle and bottom panels, respectively.
The distribution is given for an age of 6 Myr.
The histograms show the distribution of 2000 Monte Carlo simulations, while the
solid lines are based on the Edgeworth approximation. The
vertical dashed line is the average value. {\tt geneva05} stellar models and
{\tt LC2006} supernova yields were assumed.
}
\label{fig:edge}
\end{figure}

\begin{figure}
\resizebox{\hsize}{!}{\includegraphics[angle=0]{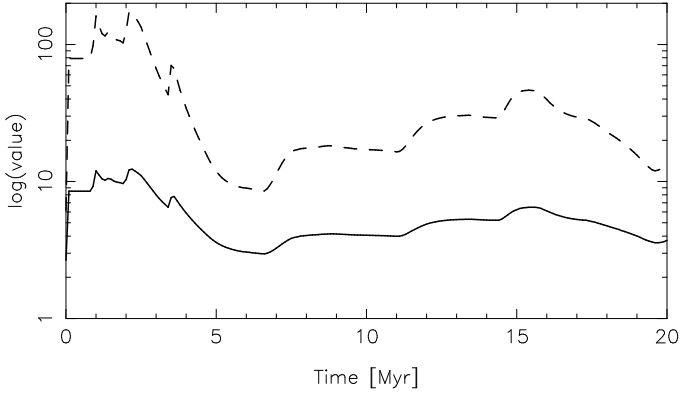}}
\caption{The skewness, $\gamma_1$ and the kurtosis, $\gamma_2$ of the
$^{26}$Al distribution for a population
of stars in the $8-120M_{\odot}$ mass range.}
\label{fig:skew}
\end{figure}

\begin{figure}
\resizebox{\hsize}{!}{\includegraphics[angle=0]{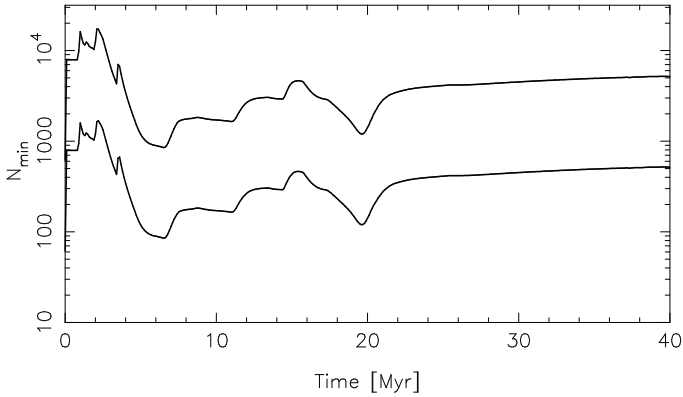}}
\caption{Lower limits on the number of stars in the $8-120 M_{\odot}$ range needed 
for the statistical distribution of $^{26}$Al to
be approximated by a Gaussian (upper line) and by Edgeworth's approximation
(lower line). {\tt geneva05} stellar models and
{\tt LC2006} supernova yields were assumed.
}
\label{fig:min}
\end{figure}

Above we described the average outputs of the $^{26}$Al, $^{60}$Fe, energy,
mass and UV radiation from young stellar populations, together with examples
of their variability, based on Monte Carlo simulations. For a better understanding
of the statistical properties of the stellar distributions, we also made
use of the statistical methods of \citet{Cervino2006}. At a given time, these 
methods allow the analytical approximation of the probability distribution 
of a quantity. We calculate the raw moments of the distribution
\begin{equation}
\mu_n=\int_{0}^{\infty}(l-a)^n\phi(l)dl,
\end{equation} 
where $n$ is the integer order of the moment, $l$ is the physical quantity for which
we investigate the distribution, and $\phi(l)$ is the probability for a specific value
of $l$. The parameter $a$ is the average value of $l$:
\begin{equation}
a=\int_{0}^{\infty}l\phi(l)dl.
\end{equation}
From these we find the skewness
\begin{equation}
\gamma_1=\frac{\mu_3}{\mu_2^{3/2}}
\end{equation}
and the kurtosis
\begin{equation}
\gamma_2=\frac{\mu_4}{\mu_2^2}-3
\end{equation}
of the distribution. For an ensemble of $N$ stars, the distribution expressed in normal
form (i.e., transformed to a distribution with zero mean and unit variance) has a skewness
$\Gamma_1=\gamma_1/\sqrt{N}$ and kurtosis $\Gamma_2=\gamma_2/N$.
With these we can estimate the distribution using
a Gaussian multiplied by Edgeworth's series \citep[see e.g.][]{Blinnikov,Cervino2006},
which, truncated to 2 terms, is:
\begin{eqnarray}
 \varphi_{\mathrm{L_{tot}}}(x) &= &\frac{1}{\sqrt{2\pi}} e^{-\frac{1}{2} x^2} \times \nonumber \\
  && \biggr( 1 + \frac{1}{6}\, \Gamma_1 \, (x^3 -3 x) + \frac{1}{24}\,\Gamma_2 \,(x^4 - 6x^2 +3)+ \nonumber \\
  && \frac{1}{72}\, \Gamma_1^2 \,(x^6-15 x^4 +45 x^2 -15)\biggl).\label{eq:edge}
\end{eqnarray}
For very small populations the discrepancy between the Edgeworth approximation
and the real distribution becomes large. \citet{Cervino2006} therefore derived
analytical estimates of the necessary population size to allow the use of their
approximation, and also when it is possible to use a Gaussian approximation.
In Fig. \ref{fig:edge} an example of the probability distribution for $^{26}$Al
is shown, and the Edgeworth approximation is compared to Monte Carlo simulations.
It can be seen that while for very low numbers of stars, the distribution can
only be investigated using Monte Carlo simulations, whereas for intermediate cases
Edgeworth's approximation is useful. Fig. \ref{fig:skew} shows the skewness and the 
kurtosis of the distribution. In Fig. \ref{fig:min} the minimum number of stars
needed for an adequate approximation (within 10\%) in the $\pm3\sigma$ interval
is shown for Egdeworths approximation and for the Gaussian approximation. 
Evidently star-forming regions need to have a very large number of  
massive stars (in the best case 1000 stars above 8 $M_\odot$) for  
Gaussian statistics to be applicable. But even outside the Gaussian  
regime, the evolution of $\gamma_1$ and $\gamma_2$ values allow us to  
obtain some inferences about the results of the Monte Carlo simulations  
with low numbers of stars without performing them. First of all, $ 
\gamma_1$ is a measure of the asymmetry of the distribution. A  
positive $\gamma_1$ means a "L-shape" distribution. A large $\gamma_2$  
means that the distribution is both more peaked than Gaussian  
distributions near the maximum, and more flat than Gaussian in the  
tails. Large $\gamma_1$ and $\gamma_2$ values together reflects then  
(a) the most probable value is peaked in a value different from the  
mean and, (b) there is a non-negligible high luminosity tail in the  
distribution.

Appling these concepts to our study, the large values in $\gamma_1$ and  
$\gamma_2$ at young ages (before the onset of the SN phase in  
$^{26}$Al production) shows the large impact of particular high  
massive stars in the the integrated production of $^{26}$Al. It  
implies that, for low mass clusters, the current $^{26}$Al produced by  
an individual cluster is strongly dependent on the most massive star  
the cluster has formed, so (provided the cluster age is known
from an independent method) 
it would be possible to estimate the value of this most  
massive star. Of course, since we are only able to obtain statistical  
inferences in the low cluster mass regime, we need to take into  
account {\it the maximum possible M$_{up}$ value}, which naturally  
include the case where such more massive stars are not present in the  
cluster due to the particular realization of the IMF performed by  
Nature in a given cluster. These effects are clearly visible even for  
cluster with 10 massive stars in the top pannel of Fig. 9. In the pure SN  
dominated phase (ages larger than 20 Myr), $\gamma_1$ and $\gamma_2$  
are almost constant, without significant structure and are both slowly  
increasing with age (cf. Fig \ref{fig:min}). A simple consequence 
of the slow decay of the SN rate at these ages is, (a) an  
increase of sampling effects due to the declining number of possible
SNe within a given time window, and (b) a decrease in the sensitivity to 
which star has exploded as a SN in the $^{26}$Al ejection, as there is not
much variation in the $^{26}$Al ejection from SNe at these late times.
The intermediate age  
range between 7 an 20 Myr shows how the stars that explode in this age  
range have different amounts of $^{26}$Al ejections.

\begin{figure}
\resizebox{\hsize}{!}{\includegraphics[angle=0]{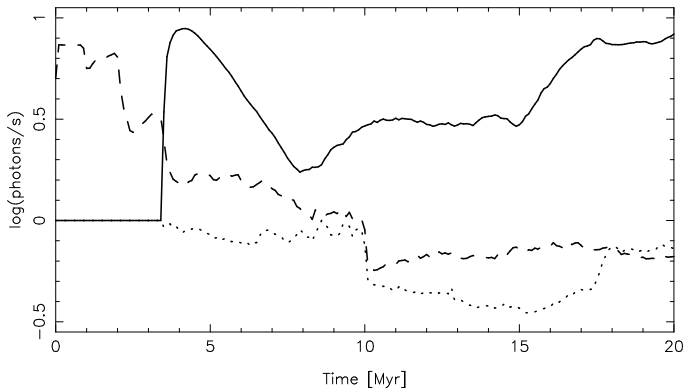}}
\caption{The Pearson correlation coefficient between the $^{26}$Al and 
$^{60}$Fe  $\gamma$-ray emission (solid
line), the $^{26}$Al $\gamma$-ray and the UV emission (dashed line) and 
the $^{60}$Fe $\gamma$-ray and the UV emission
(dotted line), as a function of time for our default model.}
\label{fig:correlation}
\end{figure}

Finally, in Fig. \ref{fig:correlation} 
the Pearson correlation coefficient between the $ 
\gamma$-ray emission from nucleosynthesis products and the hydrogen  
ionizing radiation is shown. The use of these correlation  
coefficients allows us to establish the probability of obtaining  
the value of one of the observables, given that the value of the other  
observable is known. A correlation coefficient equal to zero implies  
that the two associated quantities do not contain information about
each other. There is no strong correlation between the $\gamma$-rays from  
$^{60}$Fe and UV radiation, and therefore the overall UV  
radiation does not know anything about the $^{60}$Fe production. The  
strong correlation between the $^{26}$Al $\gamma$-ray emission  
(from the $^{26}$Al ejected in the stellar winds) and the UV radiation  
at young ages refects that both components are produced by related  
sources: Note that there is a time delay between the UV and the  
$^{26}$Al emission, so the sources that produce both quantities are  
not the same. The $^{26}$Al-UV correlation decreases with time, 
since the stars that produce the UV emission are unrelated to the stars  
that explode as SN and produce most of the $^{26}$Al at ages later
than 6 Myr. Finally, the $\gamma$-rays from $^{60}$Fe and $^{26}$Al 
are highly correlated but the correlation does not reach a value of  
1, reflecting the fact that the {\it overall} production comes from  
similar, but not the same, stellar sub-populations.

\section{Discussion}
\label{sect:discussion}

\begin{figure}
\resizebox{\hsize}{!}{\includegraphics[angle=0]{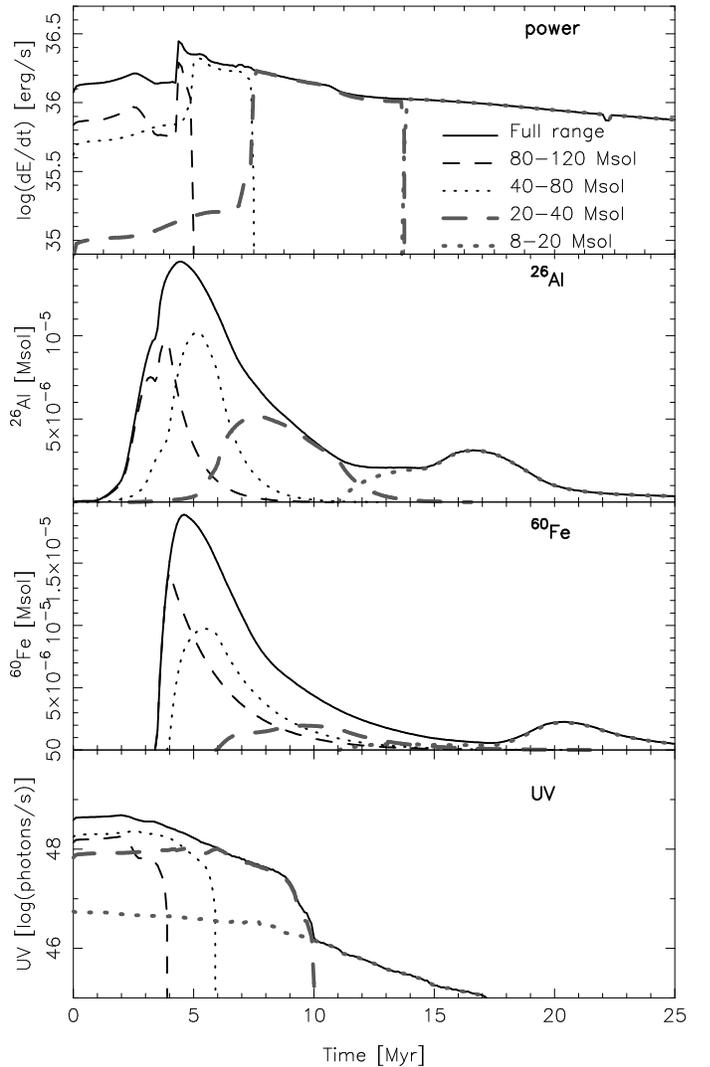}}
\caption{The time profiles of the kinetic power,
$^{26}$Al, $^{60}$Fe and UV, with the contributions from 4 mass ranges. 
The solid lines are for the full range of masses
(same as the solid lines in Figs. \ref{fig:al_timeprof} and \ref{fig:energy}.}
\label{fig:comparemasses}
\end{figure}

The outflow (energy and matter) from young stars into the ISM determines 
the interplay between star formation and galaxy evolution. Our population
synthesis code predicts important stellar outputs necessary to study these 
processes in more detail.
The emission of radioactive isotopes is important for such studies, as they
are a unique way of tracing the output from the massive stars directly, 
globally (Milky Way wide) as well
as the spatial distribution of the ejecta around young stellar clusters.
Many aspects of stellar evolution of massive stars are quite uncertain, and
often the effects of these on the properties of a stellar cluster are not directly
apparent. Population synthesis studies allow to test the impact of changes
in specific ingredients and processes of stellar evolution models.\\
Our study of the ejection of the radioactive elements $^{26}$Al and $^{60}$Fe shows
considerable theoretical uncertainties on the ejection rates and time
profiles. Especially the
amount of $^{60}$Fe ejected from the most massive stars is sensitive to
the structure of the stars in the final evolutionary stages. The emission of $^{26}$Al
varies by less than a factor $\sim$2, and the ratio of these two elements is therefore
an important diagnostic for the late stages of stellar evolution of very massive stars.
An important ingredient is mass loss from the massive stars. Despite recently improved
understanding of this process, the rate of mass loss is still subject of debate,
with proposed downward changes of up to an order of magnitude. One effect of lower
mass loss rates is an increase of the stellar core sizes at late burning stages. The
$^{60}$Fe production is correspondingly enhanced, and the wind ejection of
$^{26}$Al is decreased. For non-rotating stellar models, with the currently preferred 
mass-loss rates \citep{Limongi} the Galactic $^{60}$Fe/$^{26}$Al ratio
is overpredicted with respect to measurements \citep{Wang}, leading \citet{Limongi} to 
suggest changes in the IMF or a relatively low upper integration limit. For lower mass-loss 
rates this problem would be enhanced and this is a
strong argument against very low mass-loss rates for non-rotating stellar tracks
(although it is important to note that the uncertainties in the nucleosynthesis of $^{60}$Fe
and $^{26}$Al are too large for the argument to be definitive at this time). Stellar 
rotation can strongly enhance the mass lost from stars, mainly from increasing the duration 
of the WR-phases. However, rotation also tends to increase the size of the stellar core, and 
without calculations of the later evolutionary stages of such stars, it is difficult to predict
the explosively ejected nucleosynthetic yields. The observed average Galactic $^{60}$Fe/$^{26}$Al 
ratio provides interesting constraints on the models, but observations of specific regions are 
also important, as the time profile and the statistical variability can improve our
understanding of massive star evolution considerably. While the study of the ratio
in individual star-forming regions will not be possible for many years, it is expected
that it will at least be possible to divide the Milky Way into a few sections for this purpose
at the end of the INTEGRAL mission. The main problem is the detection limit of the
$^{60}$Fe $\gamma$-rays. On the other hand, the prospects for studying the time profile
of $^{26}$Al are good, and in subsequent papers we will apply our population synthesis tool 
to the Galactic $^{26}$Al distribution, and to individual 
star forming regions, such as Orion. Fig. \ref{fig:comparemasses} shows that
at different intervals, the output of energy and matter from the stars is
dominated by different ranges of initial stellar masses. Therefore the
analysis of different regions with non-steady star formation can be
used to explore the physics of stars in quite narrow mass ranges.\\

$^{26}$Al and $^{60}$Fe, together with many additional isotopes, are
emitted into the ISM with very high velocities. These ejecta will move away from
the star clusters, which has been observed in the Orion region where
a part of the $^{26}$Al signal is off-set from the star clusters, and coincides
with the Eridanus bubble formed by the outflowing hot gas \citep{Diehl2002}. 
The spread of $^{26}$Al can therefore trace the mixing of young stellar ejecta 
into the ISM, a process that is relatively poorly understood but important
for proper modelling of the evolution of galaxies. Our code predicts the mass, energy
and ionizing radiation output of the stellar clusters, which are the most important
parameters for the interaction with the ISM. Our results show that
stellar winds are important for the turbulent state of the ISM. In the first $\sim$6
Myr the power from the winds clearly dominates, being as strong as the subsequent power 
injected by supernovae, even if the total energy of the supernova explosions is higher 
than the total wind energy, as they explode over a much longer timescale. For the
winds the most massive stars are very important, even if there are relatively few
of them, as they lose most of their mass through winds with very high velocities in
the WR-phases. Estimating the wind power from a typical O-type star and 
multiplying by the number of stars in a cluster will underestimate 
the wind power by more than an order of magnitude. When comparing the 
mechanically derived power from different stellar models, we find that, 
while the mass-loss rates are smaller by a factor
$\sim2-3$ in the {\tt geneva05} models compared to the {\tt geneva97} models, the
wind power is actually more dominates with the {\tt geneva05} models. On the other hand,
for the {\tt LC06} models that also have the lower mass-loss rates, the role
of the winds is somewhat diminished. The reason for the enhanced wind importance when using
the rotating {\tt geneva05} models is caused by several factors, the most important
ones being the greatly increased time spent in WR-phases where the mass-loss rate and wind
velocity is very high, and the longer stellar lifetimes causing a decrease in the supernova
rate. We note that in all models the wind power is more important than commonly recognized 
(e.g. in the study of supernova driven bubbles). 
This means that for stellar clusters, the supernova shells expand into a pre-existing
cavity, rather than in a dense star-forming ISM. For example \citet{Cho} find that
the volume of a supernova bubble is $\sim2-3$ times larger and also hotter when
the winds are taken into account, even though the wind power assumed in their study
is very low compared to our estimates.\\
The flow of wind and supernova ejecta inside cavities can be very complex \citep{Maclow},
and the propagation is likely dominated by turbulent diffusion from magnetic field
irregularities caused by the stellar winds and supernovae \citep{Balsara,Parizot}.
Part of the gas will be thermalized near the stellar association due to wind-wind
collisions or by a termination shock against the turbulent medium inside the cavity,
and this can be observed as a hot X-ray emitting plasma \citep{Townsley2003,Gudel}.
However the majority of the mass will expand into the low density cavity.
The effective propagation velocity is very uncertain, and is expected to be in the range
$100-1000$ km s$^{-1}$ \citep[for example][finds velocities of $\sim$200 km s$^{-1}$]
{Balsara2008}. $^{26}$Al and $^{60}$Fe can be important for measuring this
velocity, as their lifetimes are similar to the time it takes for the ejecta to
cross a cavity blown by a young stellar cluster. If the velocity is high, these
elements will quickly reach the wall of the cavity, and the brightness distribution will
be given by the geometry of the walls. For slower propagation the fresh ejecta will
be distributed inside the cavity, and the measured line widths will reflect the
turbulence inside the cavity directly.

\section{Conclusions}
We have developed a populations synthesis code for the study of Galactic OB associations.
The main aim of this work is the prediction of the output of energy, gas, $^{26}$Al,
$^{60}$Fe and ionizing photons of a population of massive stars. We describe 
the dependence on the input physics, especially rotating stellar models
\citep{Meynet2005,Palacios}. 
We show that energy and matter output depend
on the choice of stellar evolutionary and supernova models, 
whereas other parameters, such as the
speeds of the winds and the stellar atmospheres only play secondary roles. 
This is a problem for the correct modelling of star-forming regions, and 
also means that
comparison with observations (for example the $\gamma$-ray emission from
radioactive isotopes) can potentially
yield important constraints for stellar and supernova models.
Also the statistical variations for small numbers of stars in individual 
regions are addressed, which is important when interpreting
observations of nearby, small star-forming regions.\\
Our study shows that the ejected $^{26}$Al and wind power from a population of
massive stars is strongly enhanced by the effects of stellar rotation. Despite
the recent downward revision of the mass-loss rates from massive stars by a factor
of 2-3, the amount of $^{26}$Al and the importance of the wind power is actually
increased by including stellar rotation.\\

\begin{acknowledgements}
This research was supported by the DFG cluster of excellence 'Origin and
Structure of the Universe' (http://www.universe-cluster.de). MC is supported
by the Spanish PNAyA project through FEDER funding of project AYA207-64712.
\end{acknowledgements}

\end{document}